\documentclass[journal]{IEEEtran}

\usepackage{amsmath}
\usepackage{cite}
\usepackage{graphicx}
\usepackage{epstopdf}
\usepackage{amsfonts,amsmath,amssymb}
\usepackage{graphicx}
\usepackage{url}
\usepackage{bm}
\usepackage{bbm}
\usepackage{subfigure}
\usepackage{stfloats}
\usepackage{algorithmic}
\usepackage{algorithm}
\usepackage{color}

\newtheorem{theorem}{\textbf{Theorem}}

\newtheorem{lemma}{\textbf{Lemma}}
\newtheorem{remark}{\textbf{Remark}}

\bibliography{ref}

\begin{document}
\vspace{-10mm}
\title{{Secure Transmission to the Strong User with Optimal Power Allocation in NOMA}}
\vspace{-4mm}

\author{\hspace{14mm}Youhong Feng, \IEEEmembership{Student Member, IEEE,} Shihao Yan, \IEEEmembership{Member, IEEE,} and Zhen Yang
\vspace{-12mm}
\thanks{This work was  partially supported by the National Natural Science Foundation of China (No.61772287, 61671252). }  
\thanks{Y. Feng and Z. Yang are with the Key Laboratory of Ministry of Education
 in Broadband Wireless Communication and Sensor Network Technology, Nanjing University of Posts and Telecommunications, Nanjing 210003, China (Emails: \{2013010213, yangz\}@njupt.edu.cn).  S. Yan is with the School of Engineering,  Macquarie University, Sydney, NSW, Australia (Email:shihao.yan@mq.edu.au).}}

\maketitle
\vspace{-25mm}

\begin{abstract}
With non-orthogonal multiple access (NOMA), we tackle the maximization of the secrecy rate for the strong user subject to a maximum allowable secrecy outage probability, while guaranteeing a constraint on the transmission rate to the weak user. For the first time, the dependence between the eavesdropper's ability to conduct successive interference cancellation and her channel quality is considered. We determine the optimal power allocation and the redundancy rate, based on which the cost of security in terms of the reduction in the strong user's secrecy rate is examined and the benefits of NOMA for secure transmissions are explicitly revealed.
\end{abstract}

\vspace{-1mm}

\IEEEpeerreviewmaketitle
\begin{IEEEkeywords}
Non-orthogonal multiple access, physical layer security, power allocation, redundancy rate.
\end{IEEEkeywords}
\vspace{-3mm}
\section{Introduction}

Non-orthogonal  multiple access (NOMA), which can significantly boost spectral efficiency, is envisaged as a potentially promising technique for the fifth-generation (5G) and beyond wireless communication networks\cite{L.Dai2015,Z.Ding2017JSAC}. Different from the conventional orthogonal multiple access (OMA), NOMA efficiently exploits power domain multiplexing at a transmitter and successive interference cancellation (SIC) at a receiver to serve multiple users in the same resource block (e.g., time/frequency/code domain). Specifically, these users are normally divided into two types\cite{Z.Ding2017,Z.Ding2016}, i.e., the strong users and the weak users, where the strong users require a high date rate, e.g., to support live sports streaming, while the weak users may only require a predetermined low data rate, e.g., to support text messaging.

Meanwhile, physical layer security (PLS), as a complementary and alternative technique to the traditional cryptographic methods, can defend against eavesdroppers (Eves) by exploiting the property (e.g., randomness) of the wireless medium \cite{Y. Liu2015wsl,zou2013,Y. Feng201702}. Naturally, PLS can be applied to NOMA communication networks in order to achieve the ever-lasting and information-theoretic security\cite{Y.Zhang2016,M.Tian2017sp,M. Jiang2017sp,Y.Liu2017,B.He2017}. For example,  the authors of \cite{Y.Zhang2016,M.Tian2017sp,M. Jiang2017sp} examined PLS of the wireless NOMA network with perfect knowledge of Eves'  channel state information (CSI),  which indicated a significant secrecy performance improvement achieved by NOMA relative to OMA. In addition, the authors of \cite{Y.Liu2017,B.He2017} considered a practical scenario where the transmitter does not know Eve's instantaneous CSI, where PLS of both the strong and weak users were studied.

In a NOMA system, the transmitted signals to a weak user should be decoded by the strong user in order to enable SIC at the strong user. This leads to the fact that the achieved security of the weak user is conditioned on the assumption that the strong user will not release any information transmitted to the weak user. If there is no trustiness between the weak and strong users, the information-theoretic security of weak user's information is hard to be guaranteed in NOMA systems.
Against this background, in this work we propose a new framework to examine PLS in a NOMA communication system by considering a passive eavesdropping scenario without Eve's instantaneous CSI. This framework aims to maximize the secrecy rate for the strong user subject to a maximum allowable secrecy outage probability (SOP) while guaranteeing a specific requirement on the transmission rate to the weak user. One specific applicable scenario of our new framework is where the strong user desires a high data rate with security requirement to support a tele-medical treatment, while the weak user only needs a non-secure broadcast service.

Adopting this new framework, we determine the exactly and asymptotically optimal power allocation between the strong and weak users and the optimal redundancy rate for the strong user, based on which the cost of the considered security for the strong user is explicitly examined in terms of the reduction in the secrecy rate to the strong user.
Our analysis indicates that the asymptotic results accurately approach to the corresponding exact results in the high main-to-eavesdropper ratio (MER) regime. In addition, our analysis reveals that the cost of the considered security decreases and the optimal power allocation becomes less sensitive to the maximum allowable SOP as the transmission rate requirement to the weak user increases. Furthermore, our analysis demonstrates that the equal power allocation is asymptotically optimal and the redundancy rate approaches one as the transmit power increases to infinity.
\vspace{-5mm}
\section{System Model}
\subsection{Considered Scenario and Adopted Assumptions}
We consider a NOMA communication scenario, where a transmitter  is serving two legitimate  receivers in the presence of an Eve.
We assume that each of the transceivers is equipped with a single antenna. The channel gains from the transmitter to the legitimate receivers and  Eve are denoted  $h_{k},~ k\in\{1,2\}$ and $h_{e}$, respectively,   which are the Rayleigh fading gains with independent and identically distributed (i.i.d.) entries with zero mean and variance $\delta_{k}^{2}$ and  $\delta_{e}^{2}$ respectively. We assume that CSI of all the legitimate channels is known at the transmitter, while only the statistical CSI of Eve's channel is available. Without loss of generality, we also assume that the legitimate channel gains are sorted in ascending order\cite{Y.Zhang2016,B.He2017}, i.e., $0\!\!<\!\!|h_{1}|^{2}\!\!\leq\!\! |h_{2}|^{2}$.

Employing the NOMA scheme, the transmitter sends two information signals  $s_{1}$ and $s_{2}$ to user 1 and user 2, respectively,  where the transmit power is denoted by $P$. We denote $\phi_{k}$ ($0<\phi_{k}\leq 1$) as the fraction of the transmit power allocated to user $k$ and then the transmitted superposition signal can be expressed as $\sum_{k=1}^{2}\sqrt{\phi_{k}}s_{k}$, where $\sum_{k=1}^{2}\phi_{k}=1$. In this work, we consider PLS only for the transmission to user 2 (i.e., the strong user with a higher channel gain). This is due to the fact that in a NOMA system user 2 always has to first decode the signal transmitted to user 1 (i.e., the weak user with a lower channel gain) in order to conduct SIC, which leads to the fact that the information-theoretic PLS of the signal transmitted to user 1 cannot be achieved (i.e., user 2 may release the information transmitted to user 1).
\vspace{-4mm}
\subsection{SNRs in the NOMA System}
Based on downlink NOMA scheme, user 1 decodes its own signal $s_{1}$ by treating $s_{2}$ as interference, while user 2 first decodes user 1's information (i.e., $s_{1}$) and applies SIC to decode its own information $s_{2}$. Thus,  the received signal-to-interference-plus-noise ratio (SINR) for $s_{1}$ at user 1 and signal-to-noise (SNR) for $s_{2}$  at user 2 can be expressed as
$\gamma_{1}=\frac{\phi_{1}P|h_{1}|^{2}}{\phi_{2}P|h_{1}|^{2}+\sigma_{1}^{2}}=\frac{\phi_{1}\rho_{1}|h_{1}|^{2}}{\phi_{2}\rho_{1}|h_{1}|^{2}+1},$ and
$\gamma_{2}=\frac{\phi_{2}P|h_{2}|^{2}}{\sigma_{2}^{2}}=\phi_{2}\rho_{2}|h_{2}|^{2},$
respectively, where $\rho_{1}=\frac{P}{\sigma_{1}^{2}}$ and $\rho_{2}=\frac{P}{\sigma_{2}^{2}}$.

Considering that $h_1$ and $h_2$ are known at the transmitter in the considered NOMA system, the codeword rates for $s_1$ and $s_2$ are chosen such that $R_1 \!=\! C_1$ and $R_2 \!= \!C_2$, respectively, where $C_1 \!=\!\log_{2}(1\!+\!\gamma_{1})$ and $C_2 \!=\! \log_{2}(1\!+\!\gamma_{2})$. As such, Eve can decode $s_1$ and then cancel the interference caused by $s_1$ only when Eve's channel quality is no lower than that of user 1. Therefore, there are two cases with regard to Eve's ability to decode $s_2$.
When $\rho_e|h_{e}|^{2}\!<\!\rho_1|h_{1}|^{2}$ (i.e., when Eve's channel quality is lower than that of user 1), Eve cannot decode $s_1$ and thus decodes $s_2$ directly by treating $s_1$ as interference.
In this case, the SINR of $s_{2}$ at Eve is given by $\gamma_{e1}\!=\!\frac{\phi_{2}P|h_{e}|^{2}}{\phi_{1}P|h_{e}|^{2}\!+\!\sigma_{e}^{2}}\!=\!\frac{\phi_{2}\rho_{e}|h_{e}|^{2}}{\phi_{1}\rho_{e}|h_{e}|^{2}+1}$ with $\rho_{e}\!=\!\frac{P}{\sigma_{e}^{2}}$. When $\rho_e|h_{e}|^{2}\!\geq\!\rho_1|h_{1}|^{2}$, Eve can first decode $s_1$ to in order to cancel the interference caused by $s_1$ with SIC and then decode $s_2$. Then, in this case the SNR of $s_{2}$ at Eve is given by
$\gamma_{e2}\!=\!\frac{\phi_{2}P|h_{e}|^{2}}{\sigma_{e}^{2}}\!=\!\phi_{2}\rho_{e}|h_{e}|^{2}.$
Considering Rayleigh fading, the cumulative distribution functions (cdfs) of $\gamma_{e1}$ and $\gamma_{e2}$ are obtained as
\vspace{-1mm}
\setcounter{equation}{0}
\begin{align}\label{cdf-e-1}
F_{\gamma_{e1}}(\gamma)&=\bigg\{\begin{array}{ll}
1-e^{-\frac{\gamma}{\rho_{e}(\phi_{2}-\phi_{1}\gamma)}}, & \gamma\leq\frac{\phi_{2}}{\phi_{1}},\\
1, & \gamma>\frac{\phi_{2}}{\phi_{1}},
\end{array}
\bigg.\\\label{cdf-e-2}
F_{\gamma_{e2}}(\gamma)&=1-e^{-\frac{\gamma}{\phi_2\rho_e}}.
\end{align}
\vspace{-9mm}
\section{Physical Layer Security of the Strong User}
In this section, we aim to maximize the secrecy rate subjective to a maximum allowable SOP for user 2 while guaranteeing a requirement on the transmission rate to user 1. The exactly and asymptotically optimal power allocation coefficients $\phi_{k}$ and the redundancy rate are determined, based on which the associated maximum secrecy rate is determined.
\vspace{-9mm}
\subsection{Transmission Scheme and the Optimization Problem}
In order to achieve PLS for user 2, in addition to $R_2$, a redundancy rate $R_E$ for transmitting $s_2$ should be determined \cite{S.Yan2017TVT}. Then, the instantaneous secrecy rate for user 2 as a function of $\phi_{2}$ and $R_{E}$ is given by
\begin{align}\label{RS-2}
R_{2}^{s}(\phi_{2},R_{E})\! =\! [C_{2}-R_{E}]^{+}= [\log_{2}(1\!+\!\phi_{2}\tilde{\rho_{2}})\!-\!R_{E}]^{+},
\end{align}
where $\tilde{\rho_{2}}={\rho_{2}}|h_{2}|^{2}$ and $[x]^{+}=\max(x,0)$. However, a secrecy outage occurs when $C_{E}>R_{E}$, where $C_{E}=\log_{2}(1+\gamma_{e}),~\gamma_{e}\in\{\gamma_{e1},\gamma_{e2}\}$ is the unknown channel capacity of Eve. We derive the SOP in the following lemma.
\vspace{0mm}
\begin{lemma}\label{Lemma 1}
For our considered NOMA communication scenario, the SOP of $s_2$ is given by
\begin{align}\label{YOHU-9}
&P_{out}(R_{E},\phi_{2}) \!\!= \!\!\notag\\
&(1\!-\!e^{\!-\!\frac{\rho_1|h_1|^{2}}{\rho_e}})P_{out,1}(R_{E},\phi_{2})+e^{-\frac{\rho_1|h_1|^{2}}{\rho_e}}P_{out,2}(R_{E},\phi_{2}),\end{align}
where $P_{out,1}(R_{E},\phi_{2})=e^{-\frac{2^{R_{E}}-1}{\rho_{e}(\phi_{2}-\phi_{1}(2^{R_{E}}-1))}}$ when $2^{R_{E}}-1<\frac{\phi_{2}}{\phi_{1}}$, $P_{out,1}(R_{E},\phi_{2})=0$ when $2^{R_{E}}-1\geq\frac{\phi_{2}}{\phi_{1}}$, and $P_{out,2}(R_{E},\phi_{2})=e^{-\frac{2^{R_{E}}-1}{\rho_{e}\phi_{2}}}$.
\end{lemma}
\begin{IEEEproof}
Based on the analysis in Section II-B, the SOP of $s_2$ can be given by
\begin{align}\label{YOHU-9-0}
P_{out}(R_{E},\phi_{2}) &\!=\!Pr(\rho_e|h_{e}|^{2}\!<\!\rho_1|h_{1}|^{2})P_{out,1}(R_{E},\phi_{2})\notag\\
&\!+\!Pr(\rho_e|h_{e}|^{2}\!\geq\!\rho_1|h_{1}|^{2})P_{out,2}(R_{E},\phi_{2}),
\end{align}
where
$P_{out,1}(R_{E},\phi_{2}) = Pr(\gamma_{e1} \geq 2^{R_E} - 1)$ and $P_{out,2}(R_{E},\phi_{2}) = Pr(\gamma_{e2} \geq 2^{R_E} - 1)$. Considering Rayleigh fading for $h_e$, we have $Pr(\rho_e|h_{e}|^{2}\!<\!\rho_1|h_{1}|^{2})\!=\!1-e^{-\frac{\rho_1|h_1|^{2}}{\rho_e}}$ and $Pr(\rho_e|h_{e}|^{2}\!>\!\rho_1|h_{1}|^{2})\!=\!e^{-\frac{\rho_1|h_1|^{2}}{\rho_e}}$. Then, substituting \eqref{cdf-e-1} and \eqref{cdf-e-2} into \eqref{YOHU-9-0}, we obtain \eqref{YOHU-9}, which proves Lemma~\ref{Lemma 1}.
\end{IEEEproof}
\begin{remark}\label{remark0}
Following Lemma~1, we note that there is a minimum value for the SOP in order to guarantee a positive secrecy rate, since $h_e$ is unknown and the maximum value of $R_E$ is $C_2$ to ensure a positive secrecy rate as per \eqref{RS-2}. We note that this minimum value of the SOP exists in both the NOMA and OMA systems. In the NOMA system, this minimum value is $\varepsilon_n\!=\! e^{\!-\!\frac{\sigma_e^2|h_1|^{2}}{\sigma_1^2}}e^{-\frac{\sigma_e^2 |h_2|^2}{\sigma_2^2}}$, which is achieved by substituting $R_E \!= \!C_2$ into \eqref{YOHU-9} and noting $P_{out,1}(R_{E},\phi_{2})$ can be enforced to zero by varying $\phi_2$. In the OMA system, this minimum value is $\varepsilon_o\!\!=\!\! e^{\!-\!\frac{\sigma_e^2 |h_2|^2}{\sigma_2^2}}$, since in the OMA system Eve does not have interference for decoding $s_2$. We note that $\varepsilon_n \!\!< \!\!\varepsilon_o$ due to $e^{\!-\!\frac{\sigma_e^2|h_1|^{2}}{\sigma_1^2}}\!\! <\!\! 1$, which is one explicit benefit of NOMA for secure transmission. Specifically, for $\varepsilon_n \!\!\leq\!\! \varepsilon\! \!<\! \!\varepsilon_o$, where $\varepsilon$ is the maximum allowable SOP for user 2 in the considered system, NOMA can achieve a positive secrecy rate, while OMA cannot. In this work we assume $\varepsilon_n\! \!\leq\! \!\varepsilon$ in order to guarantee the feasibility of the considered optimization problem and in the comparison between NOMA and OMA we consider $\varepsilon_o \!\leq \!\varepsilon$.
\end{remark}

In this work, we optimize the power allocation coefficients (i.e., $\phi_{1}$ and $\phi_{2}$) and the redundancy rate $R_{E}$ to maximize the instantaneous secrecy rate $R_{2}^{s}(\phi_{2},R_{E})$ subject to the maximum allowable SOP $\varepsilon$ for user 2 and a minimum data rate $Q_1$ for user 1. Then, the optimization problem at the transmitter is given by
\begin{align}\label{YOHU-3}
\textbf{P1}:&\operatorname*{max}\limits_{R_{E},\phi_{1},\phi_{2}} R_{2}^{s}(\phi_{2},R_{E})\,\,\\\label{YOHU-3a}
&{\rm  s.t.}\hspace{3mm}C_1\geq Q_{1}, \\ \label{YOHU-3b}
&\hspace{7.5mm} P_{out}(R_{E},\phi_{2}) \leq \varepsilon,\\ \label{YOHU-3c}
&\hspace{8mm} \phi_{1}+\phi_{2}=1.
\end{align}

\begin{lemma}\label{Lemma 2}
For any given $\phi_{1}$ or $\phi_{2}$, there is a unique $R_{E}$ that maximizes $R_{2}^{s}(\phi_{2},R_{E})$ subject to $ P_{out}(R_{E},\phi_{2}) \leq \varepsilon$ and
this value is $R_E^{\dag}(\phi_2)$ that guarantees $P_{out}(R_{E},\phi_{2}) = \varepsilon$.
\end{lemma}
\begin{IEEEproof}
Following \eqref{RS-2}, we note that $R_{2}^{s}(\phi_{2},R_{E})$ monotonically decreases with $R_E$, while $P_{out}(R_{E},\phi_{2})$ is a  non-increasing function of $R_{E}$ as per \eqref{YOHU-9}. This proves Lemma~\ref{Lemma 2}.\end{IEEEproof}
\vspace{-3mm}
\subsection{Exact Solution to the Optimization Problem $\textbf{P1}$}\begin{theorem}\label{Theorem 1}
The feasible condition of \textbf{P1} is $P>\max(\frac{2^{Q_{1}}-1}{|h_{1}|^{2}}\sigma_{1}^{2}, P_{min} )$, under which the optimal power allocation coefficients and redundancy rate are derived as
 \begin{align}\label{Oprimal-coff-3-0}
&\phi_{2}^{\ast}\!=\!\min\!\bigg(\!\underbrace{\frac{1}{2^{Q_1}}\!+\!\frac{\sigma_1^2}{P|h_{1}|^{2}2^{Q_{1}}} \!-\! \frac{\sigma_1^2}{P|h_{1}|^{2}}}_{\phi_2^{\dag}},~~\phi_2^{\ddag}\!\bigg),\!\\
&\phi_{1}^{\ast}=1-\phi_{2}^{*}, \label{optimal_phi01}\\ \label{Oprimal-coff-3-1}
&R_{E}^{\ast}=R_E^{\dag}(\phi_2^{\ast}),
\end{align}
where $\phi_2^{\ddag}$ is the value of $\phi_2$ that maximizes $\log_{2}(1+\phi_{2}\rho_{2}|h_{2}|^{2})-R_E^{\dag}(\phi_2)$.
\end{theorem}\begin{IEEEproof}
 Following the expression of $\gamma_{1}$, $C_1$ is maximized when $\phi_1 \rightarrow 1$ and $\phi_2 \rightarrow 0$. In order to guarantee the constraint \eqref{YOHU-3a}, we have to ensure $\rho_1 \geq \frac{2^{Q_{1}}-1}{|h_{1}|^{2}}$, which leads to $P\geq\frac{2^{Q_{1}}-1}{|h_{1}|^{2}}\sigma_{1}^{2}$.
As per Lemma~\ref{Lemma 2}, we have $P_{out}(R_{E},\phi_{2}) = \varepsilon$ in the solution to \textbf{P1}, which leads to $R_E = R_E^{\dag}(\phi_2)$.  Substituting this value of $R_E$ into \eqref{RS-2}, we have $R_2^{s}(\phi_{2})=\log_{2}(1+\phi_{2}\rho_{2}|h_{2}|^{2})-R_E^{\dag}(\phi_2).$
In order to guarantee $R_2^{s}(\phi_{2}) \! \!> \! \! 0$, we have to ensure $\phi_{2}\rho_{2}|h_{2}|^{2}  \! \!>  \! \!2^{R_E^{\dag}(\phi_2)} \! \!- \! \!1$,  which requires that $P_{min}$ satisfies $\phi_{2}\rho_{2}|h_{2}|^{2}  \! \!>  \! \!2^{Q(\phi_2)} \! \!- \! \!1$ with the constraint of $\phi_1 \!+ \!\phi_2 \!= \!1$.
Under this feasible condition, the optimal $\phi_{2}$ that maximizes $R_2^{s}(\phi_{2})$ without the constraint $C_1 \geq Q_1$ can be obtained through $\phi_2^{\ddag}\!\!=\!\!\operatorname*{argmax}\limits_{\phi_2}R_2^{s}(\phi_{2})\!\!=\!\!\operatorname*{argmax}\limits_{\phi_2}(\log_{2}(1\!+\!\phi_{2}\rho_{2}|h_{2}|^{2})\!\!-\!\!R_E^{\dag}(\phi_2)).$
By setting $C_1 \!=\! Q_1$ in \eqref{YOHU-3a}, we have $\phi_2\! =\! \frac{\rho_{1}|h_{1}|^{2}-2^{Q_{1}}\!+\!1}{\rho_{1}|h_{1}|^{2}2^{Q_{1}}}$. Due to $\phi_1\! + \!\phi_2\! =\! 1$, $C_1$ is a monotonically decreasing function of $\phi_2$ and thus in order to guarantee $C_1 \geq Q_1$ we have the desired result in \eqref{Oprimal-coff-3-0}, based on which \eqref{optimal_phi01} and \eqref{Oprimal-coff-3-1} are achieved as per $\phi_1 \!+ \!\phi_2 \!= \!1$ and Lemma~2, respectively.
\end{IEEEproof}\begin{remark}\label{remark2}
Comparing $\varepsilon_n\!\leq\!\varepsilon\!\!<\!\!1$ in Theorem~1 with $\varepsilon\!\!=\!\!1$ (i.e., the optimization problem in \eqref{YOHU-3} without  security constraint in \eqref{YOHU-3b}), we note that the considered PLS is not only at the cost of a positive redundancy rate $R_E^{\ast}$, but also a possible reduction in $\phi_2^{\ast}$, since this $\phi_2^{\ast}$ is no larger than that in case without considering security. Meanwhile, as per \eqref{Oprimal-coff-3-0} in Theorem~1, we note that $\phi_2^{\dag}$ decreases with $Q_1$. This indicates that, as $Q_1$ increases, the power allocation will be more likely determined by $C_1 \!\!\geq \!\!Q_1$ and not sensitive to the required secrecy levels (i.e., the values of $\varepsilon$). We also note that when $\phi_2^{\ast} \!=\! \phi_2^{\dag}$ the cost of PLS in terms of $R_E^{\ast}$ decreases with $Q_1$. Intuitively, this is due to the fact that the power allocated to user 1 also causes interference at Eve.
\end{remark}
\vspace{-4.5mm}
\subsection{Asymptotic Solution to the Optimization Problem $\textbf{P1}$ }
In order to provide useful insights on the secrecy performance of our considered communication system, in this subsection we focus on the secrecy rate $R_{2}^{s}(\phi_{2},R_{E})$ with a high MER (i.e., ${\rho_{1}}/{\rho_{e}}=\sigma_e^2/\sigma_1^2\rightarrow \infty$) \cite{zou2013}.
\begin{lemma}\label{Lemma 3}
With ${\rho_{1}}/{\rho_{e}}\rightarrow \infty$, the asymptotic SOP of $s_2$ is derived as
\vspace{-1mm}
\begin{align}\label{POUT-ASY-s2}
\hspace{-3mm}P_{out}^{\infty}(R_{E},\phi_{2})\!\!=\!\!\left\{\begin{array}{ll}
e^{-\frac{2^{R_{E}}\!-\!1}{\rho_{e}\left(\phi_{2}-\phi_{1}(2^{R_{E}}\!-\!1)\right)}}, & 2^{R_{E}}\!-\!1\!<\!\frac{\phi_{2}}{\phi_{1}},\\
0, & 2^{R_{E}}\!-\!1\!\geq\!\frac{\phi_{2}}{\phi_{1}}.
\end{array}
\right.
\end{align}
\end{lemma}
\begin{IEEEproof}
Following \eqref{YOHU-9},  we have $\lim_{\rho_{1}/\rho_{e}\rightarrow \infty}e^{-\frac{\rho_1|h_1|^{2}}{\rho_e}}\!=\!0$.   As such, we conclude that the asymptotic SOP of $s_2$ is equal to $P_{out,1}(R_{E},\phi_{2})$ when $\frac{\rho_{1}}{\rho_{e}}\!\rightarrow\! \infty$. This completes the proof.\end{IEEEproof}
\begin{theorem}\label{Theorem 2}
For ${\rho_{1}}/{\rho_{e}}\rightarrow \infty$ with $\varepsilon>0$, $P>\max\left(\frac{2^{Q_{1}}-1}{|h_{1}|^{2}}\sigma_{1}^{2}, \frac{\sigma_{2}^{2}}{|h_{2}|^{2}}\!-\!\frac{\sigma_e^2}{\ln(\frac{1}{\varepsilon})}\right)$ is the feasible condition of \textbf{P1}, under which the optimal power allocation coefficients and redundancy rate are derived as
\begin{align}\label{Oprimal-coff-3-00}
&\phi_{2}^{\ast}\!=\!\!\notag\\
&\min\bigg(\underbrace{\frac{1}{2^{Q_1}}\!+\!\frac{\sigma_1^2}{P|h_{1}|^{2}2^{Q_{1}}} \!-\! \frac{\sigma_1^2}{P|h_{1}|^{2}}}_{\phi_2^{\dag}},
\underbrace{\frac{1}{2} \!+\! \frac{\sigma_e^2}{2 P \ln(\frac{1}{\varepsilon})} \!-\! \frac{\sigma_2^2}{2 P |h_{2}|^{2}}}_{\phi_2^{\ddag}}\bigg),\\
&\phi_{1}^{\ast}=1-\phi_{2}^{*}, \label{optimal_phi1}\\ \label{Oprimal-coff-3-01}
&R_{E}^{\ast}=\log_{2}\left(1+\frac{\phi_{2}^{*}P\ln(\frac{1}{\varepsilon})}{\sigma_e^2+\phi_{1}^{*}P\ln(\frac{1}{\varepsilon})}\right).
\end{align}
\end{theorem}
\begin{IEEEproof}
Similar to the proof of Theorem~\ref{Theorem 1}, in order to guarantee $C_1 \geq Q_1$ we have $P\geq\frac{2^{Q_{1}}-1}{|h_{1}|^{2}}\sigma_{1}^{2}$. As per Lemma~\ref{Lemma 1}, we have $P_{out}^{\infty}(R_{E},\phi_{2}) = \varepsilon$ in the solution to \textbf{P1}, which leads to $R_E = \log_{2}\left(1+\frac{\phi_{2}\rho_{e}\ln(\frac{1}{\varepsilon})}{1+(1-\phi_{2})\rho_{e}\ln(\frac{1}{\varepsilon})}\right)$ based on \eqref{YOHU-9} for $\varepsilon >0$.  Substituting this value of $R_E$ into \eqref{RS-2}, we have
$R_2^{s}(\phi_{2})=\log_{2}(1+\phi_{2}\rho_{2}|h_{2}|^{2})-\log_{2}(1+\frac{\phi_{2}\rho_{e}\ln(\frac{1}{\varepsilon})}{1+(1-\phi_{2})\rho_{e}\ln(\frac{1}{\varepsilon})}).$ In order to guarantee $R_2^{s}(\phi_{2}) > 0$, we have to ensure $\phi_{2}\rho_{2}|h_{2}|^{2} > \frac{\phi_2 \rho_{e}\ln(\frac{1}{\varepsilon})}{1+(1-\phi_{2})\rho_{e}\ln(\frac{1}{\varepsilon})},$ which at least requires
\begin{align}\label{Tiaojian-11}
\rho_{2}|h_{2}|^{2} > \frac{\rho_{e}\ln(\frac{1}{\varepsilon})}{1+\rho_{e}\ln(\frac{1}{\varepsilon})},
\end{align} since $\frac{\rho_{e}\ln(\frac{1}{\varepsilon})}{1+(1-\phi_{2})\rho_{e}\ln(\frac{1}{\varepsilon})}$ is minimized when
$\phi_2 \rightarrow 0$. We note that \eqref{Tiaojian-11} leads to $P > \frac{\sigma_e^2}{\ln(\frac{1}{\varepsilon})}\left(\frac{\sigma_2^2\ln(\frac{1}{\varepsilon})}{\sigma_{e}^{2}|h_{2}|^{2}}-1\right)$.
As such, the feasible condition of \textbf{P1} is $P>\max\left(\frac{2^{Q_{1}}-1}{|h_{1}|^{2}}\sigma_{1}^{2}, \frac{\sigma_e^2}{\ln(\frac{1}{\varepsilon})}\left(\frac{\sigma_2^2\ln(\frac{1}{\varepsilon})}{\sigma_{e}^{2}|h_{2}|^{2}}-1\right)\right)$.
Under this feasible condition, the first and second derivatives of $R_2^{s}(\phi_{2})$ with respect to $\phi_2$ are derived as
  \begin{align}\label{Tiaojian-2}
\hspace{-5mm}\frac{\partial R_2^{s}(\phi_{2})}{\partial \phi_{2}}&\!\!=\!\!\frac{1}{\ln2}\!\left(\!\frac{\rho_{2}|h_{2}|^{2}}{1\!+\!\phi_{2}\rho_{2}|h_{2}|^{2}}\!\!-\!\!\frac{\rho_{e}\ln(\frac{1}{\varepsilon})}{1\!+\!(1\!-\!\phi_{2})\rho_{e}\ln(\frac{1}{\varepsilon})}\!\right)\!,\\
\frac{\partial^{2} R_2^{s}(\phi_{2})}{\partial \phi_{2}^{2}}&=\frac{1}{\ln2}\!\!\left(\!\!\frac{-\rho_{2}^{2}|h_{2}|^{4}}{(1\!+\!\phi_{2}\rho_{2}|h_{2}|^{2})^{2}}\!\!-\!\!
\frac{(\rho_{e}\ln(\frac{1}{\varepsilon}))^{2}}{(1\!+\!(1\!-\!\phi_{2})\rho_{e}\ln(\frac{1}{\varepsilon}))^{2}}\!\!\right)\!\!.\label{Tiaojian-3}
    \end{align}
Following \eqref{Tiaojian-3}, we have ${\partial^{2} R_2^{s}(\phi_{2})}/{\partial \phi_{2}^{2}} < 0$ by noting $0<\varepsilon <1$, which indicates that $R_2^{s}(\phi_{2})$ is a concave function of $\phi_{2}$. As such, as per \eqref{Tiaojian-2} by setting ${\partial R_2^{s}(\phi_{2})}/{\partial \phi_{2}} =0$ we achieve $\phi_2 = \frac{\rho_{2}|h_{2}|^{2}(1+\rho_{e}\ln(\frac{1}{\varepsilon}))-\rho_{e}\ln(\frac{1}{\varepsilon})}{2\rho_{2}|h_{2}|^{2}\rho_{e}\ln(\frac{1}{\varepsilon})}$. By setting $C_1 = Q_1$ in \eqref{YOHU-3a}, we have $\phi_2 = \frac{\rho_{1}|h_{1}|^{2}-2^{Q_{1}}+1}{\rho_{1}|h_{1}|^{2}2^{Q_{1}}}$. Due to $\phi_1 + \phi_2 = 1$, $C_1$ is a monotonically decreasing function of $\phi_2$ in order to guarantee $C_1 \geq Q_1$. We can obtain  the desired result in \eqref{Oprimal-coff-3-00}, based on which \eqref{optimal_phi1} and \eqref{Oprimal-coff-3-01} are obtained.
\end{IEEEproof}
\begin{remark}\label{remark3}
Based on \eqref{Oprimal-coff-3-00} in Theorem~\ref{Theorem 2}, we note that as $P \rightarrow \infty$ we have $\phi_2^{\dag} \rightarrow \frac{1}{2^{Q_1}}$ and $\phi_2^{\ddag} \rightarrow \frac{1}{2}$. This indicates that in the limit of $P \rightarrow \infty$ we have $\phi_2^{\ast} = \frac{1}{2^{Q_1}}$ when $Q_1 > 1$ and $\phi_2^{\ast} = \frac{1}{2}$ when $Q_1 \leq 1$. As per \eqref{Oprimal-coff-3-01} in Theorem~\ref{Theorem 2}, we note that as $P \rightarrow \infty$ we have $R_E^{\ast} \rightarrow \log_{2}\left(1+\frac{\phi_{2}^{\ast}}{\phi_1^{\ast}}\right)$. Therefore, we can conclude that in the limit of $P \rightarrow \infty$ equal power allocation is optimal and $R_E^{\ast} = \log_2(2)$ when $Q_1 \leq 1$, and the power allocation is solely determined by the required transmission rate to user 1 while $R_E^{\ast}$ decreases with $Q_1$ when $Q_1 > 1$. In addition, we can see that in the limit of $P \rightarrow \infty$ the power allocation and the cost of PLS are regardless of $\varepsilon$. Furthermore, we note that for $\varepsilon \rightarrow 0$ we have $\phi_2^{\ast} < 1/2$ regardless of $P$, due to $\phi_2^{\ddag} < 1/2$ in Theorem~2 and thus $R_E^{\ast} \leq \log_2(2)$. This indicates that for a strict secrecy requirement, the transmitter will not allocate more power to user 2 relative to user 1.
\end{remark}
\vspace{-4mm}
\section{Numerical Results}
\begin{figure}[!t]
 \centering\
  \includegraphics[height=2.30in]{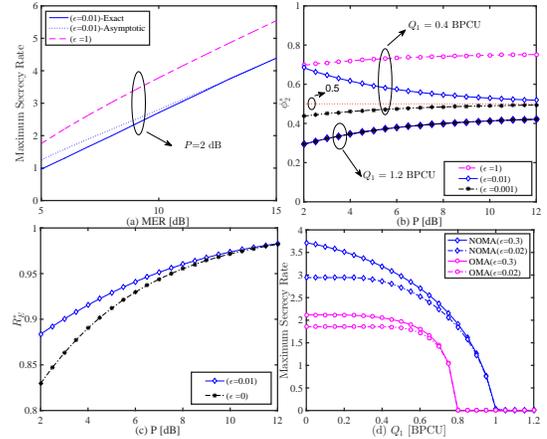}
 \caption{(a) Maximum secrecy rate versus MER,   (b) $\phi_2^{\ast}$ versus $P$,  (c) $R_E^{\ast}$ versus $P$,  and (d)  Maximum secrecy rate versus  $Q_{1}$, where $Q_1=0.4$ for (a) and (c),  $P=4$dB for (d),  $\sigma_{1}^2=\sigma_{2}^2=-5$dB and $\sigma_{e}^2=2$dB for (b)-(d), $|h_{1}|^{2}=0.5$, $|h_{2}|^{2}=4$ for (a)-(d).}
 \label{fig_sim2}
 \end{figure}

In Fig.~\ref{fig_sim2}~(a),  we plot the maximum secrecy rate for user 2 achieved by our exact and asymptotic solutions to \text{P1}. In Fig.~\ref{fig_sim2}~(b) and Fig.~\ref{fig_sim2}~(c), we plot  $\phi_2^{\ast}$ and $R_E^{\ast}$ versus the transmit power $P$, respectively.
In Fig.~\ref{fig_sim2}~(a), we first observe that the asymptotic results approach to the exact
results in the high MER regime, which verifies the correctness of our analysis. In Fig.~\ref{fig_sim2}~(a), we also observe that $R_2^{\ast}$ for $\varepsilon<1$ is lower than that for $\varepsilon=1$ (without considering security), where the performance reduction is the cost of considered PLS.
In Fig.~\ref{fig_sim2}~(b), we observe that $\phi_2^{\ast}$ for either $\varepsilon = 0.01$ or $\varepsilon = 0.001$ is lower than that for $\varepsilon = 1$, which shows one specific cost of the considered PLS that less power will be allocated to user 2. As expected from Remark~\ref{remark3}, $\phi_2^{\ast}$ for either $\varepsilon = 0.01$ or $0.001$ approaches $1/2$ as $P \rightarrow \infty$ when $Q_1$ is small, which demonstrates the optimality of the equal power allocation in the limit of $P \rightarrow \infty$ with a small $Q_1$.
In Fig.~\ref{fig_sim2}~(c), we observe that $R_E^{\ast}$ increases with $P$ and $R_E^{\ast} \rightarrow 1$ as $P \rightarrow \infty$, which can be explained by our theorems and Remark~\ref{remark3}.
In Fig.~\ref{fig_sim2}~(d), we plot the maximum secrecy rate achieved by the NOMA and OMA schemes versus $Q_1$. Here, the OMA scheme refers to the conventional TDMA scheme, in which half of the transmission time is allocated to user 1 and the other half is allocated to user 2. In this figure, we first observe that the maximum secrecy rate $R_2^{\ast}$ achieved by the NOMA scheme is significantly higher than that achieved by the OMA scheme under the specific system settings. In addition, we observe that the NOMA scheme can still achieve a positive secrecy rate when the OMA scheme cannot (i.e., when $0.8\leq Q_1 \leq 1$). This is due to the fact that the transmission time to user 1 in the OMA scheme is only half of that in the NOMA scheme and thus it may require a higher transmit power to user 1 in the OMA scheme.
\vspace{-4.5mm}
\section{Conclusion}
In this work, we determined the optimal power allocation and redundancy rate in order to maximize the secrecy rate for the strong user subject to a maximum allowable SOP, while guaranteeing the non-secure transmission rate requirement to the weak user. The security of the weak user was not considered, since the signal dedicated to the weak user should be decoded by the strong user for SIC in NOMA. Our analysis explicitly revealed the cost of the considered PLS in terms of the reduction in the secrecy rate to the strong user, which as shown decreases as the requirement on the transmission rate to the weak user increases.


\vspace{-4mm}
\begin{thebibliography}{10}
\bibitem{L.Dai2015}
L. Dai, B. Wang, Y. Yuan, S. Han, C.-L. I, and Z. Wang, ``Non-orthogonal
multiple access for 5G: Solutions, challenges, opportunities,
and future research trends,'' \emph{IEEE Commun. Mag.}, vol. 53, no. 9, pp.
74--81, Sep. 2015.
\bibitem{Z.Ding2017JSAC}
 Z. Ding, X. Lei, G. K. Karagiannidis, R. Schober, J. Yuan, and V. K. Bhargava, ``A survey on non-orthogonal multiple access for 5G
networks: Research challenges and future trends,''  \emph{IEEE J. Select. Areas Commun.}, vol. 35, no. 10, pp. 2181--2195, Oct. 2017.

 \bibitem{Z.Ding2017}
Z.  Ding,  P.  Fan,  and  H.  V.  Poor, ``Impact  of  user  pairing  on  5G
non-orthogonal  multiple-access  downlink  transmissions,''  \emph{
IEEE  Trans.
Veh. Technol.}, vol. 65, no. 8, pp. 6010--6023,  Aug. 2016.
\bibitem{Z.Ding2016}
Z. Ding, H. Dai, and H. V. Poor, ``Relay selection for cooperative
NOMA,''  \emph{IEEE Wireless Commun. Lett.}, vol. 5, no. 4, pp. 416--419, Aug. 2016.

\bibitem{Y. Liu2015wsl}
Y. Liu, L. Wang, T. T. Duy, M. Elkashlan, and T. Q. Duong, ``Relay
selection for security enhancement in cognitive relay networks,'' \emph{IEEE
Wireless Commun. Lett.}, vol. 4, no. 1, pp. 46--49, Feb. 2015.

\bibitem{zou2013} Y. Zou, X.Wang, and W. Shen, ``Optimal relay selection for physical-layer
security in cooperative wireless networks,'' \emph{IEEE J. Sel. Areas Commun.},
vol. 31, no. 10, pp. 2099--2111, Oct. 2013.

\bibitem{Y. Feng201702}
 Y. Feng, S. Yan, Z. Yang, N. Yang, and W.-P. Zhu, `TAS-based incremental hybrid decode-amplify-forward relaying for physical layer security enhancement,''  \emph{IEEE Trans. Commun.}, vol. 65, no. 9 pp. 3876--3891, Sep. 2017.

\bibitem{Y.Zhang2016}
Y. Zhang, H.-M. Wang, Q. Yang, and Z. Ding, ``Secrecy sum rate
maximization in non-orthogonal multiple access,''  \emph{
IEEE Commun. Lett.},
vol. 20, no. 5, pp. 930--933, May 2016.

\bibitem{M. Jiang2017sp}
M. Jiang, Y. Li, Q. Zhang, Q. Li, and J. Qin, ``Secure beamforming in downlink MIMO
non-orthogonal multiple access networks,''  \emph{IEEE
Signal Process. Lett.}, vol. 24, no. 12, pp. 1852--1855, Aug. 2017.
\bibitem{M.Tian2017sp}
M. Tian, Q. Zhang, S. Zhao, Q. Li, and J. Qin, ``Secrecy sum rate optimization for downlink MIMO non-orthogonal multiple access systems,''  \emph{IEEE
Signal Process. Lett.}, vol. 24, no. 8, pp. 1113--1117, Aug. 2017.

\bibitem{Y.Liu2017}
Y. Liu,  Z. Qin, M. Elkashlan,
Y. Gao, and L. Hanzo, ``Enhancing the physical layer security of
non-orthogonal multiple access in
large-scale networks,'' \emph{ IEEE Trans. Wireless Commun.}, vol. 16, no. 3, pp. 1656--1671, Dec. 2017.
\bibitem{B.He2017}
 B. He, A. Liu, N. Yang, and V. K. N. Lau, ``On the design of secure non-orthogonal
multiple access systems,''  \emph{IEEE J. Select. Areas Commun.}, vol. 35,
no. 10, pp. 2196--2206, Oct. 2018.





\bibitem{S.Yan2017TVT}
S. Yan, N. Yang, Ingmar Land, R. Malaney, and J. Yuan, ``Three artificial-noise-aided secure transmission schemes in wiretap channels,'' \emph{ IEEE Trans. Veh. Technol.}, vol. 67, no. 4, pp. 3669--3673,  Apr. 2018.



\end{thebibliography}
\end{document}